# Lane-free Artificial-Fluid Concept for Vehicular Traffic

**Markos Papageorgiou[*], Kyriakos-Simon Mountakis[*],
Iasson Karafyllis[**] and Ioannis Papamichail[*]**

[*]Dynamic Systems and Simulation Laboratory, Technical University of Crete,
Chania, 73100, Greece,
email: markos@dssl.tuc.gr ; kmountakis@dssl.tuc.gr ; ipapa@dssl.tuc.gr

[**]Dept. of Mathematics, National Technical University of Athens,
Zografou Campus, 15780, Athens, Greece,
email: iasonkar@central.ntua.gr ; iasonkaraf@gmail.com

**Abstract**

A novel paradigm for vehicular traffic in the era of connected and automated vehicles (CAVs) is proposed, which includes two combined principles: lane-free traffic and vehicle nudging, whereby vehicles are "pushing" (from a distance, using communication or sensors) other vehicles in front of them. This traffic paradigm features several advantages, including: smoother and safer driving; increase of roadway capacity; and no need for the anisotropy restriction. The proposed concept provides, for the first time since the automobile invention, the possibility to actively design (rather than describe) the traffic flow characteristics in an optimal way, i.e. to engineer the future CAV traffic flow as an efficient artificial fluid. Options, features, application domains and required research topics are discussed. Preliminary simulation results illustrate some basic features of the concept.

**Keywords:** automated vehicles, traffic flow, vehicle nudging, artificial fluid.

## 1. Introduction

Vehicular traffic has evolved as a crucial means for the transport of persons and goods, and its importance for the economic and social life of modern society cannot be overemphasized. On the other hand, vehicular traffic congestion, which appears on a daily basis, particularly in and around metropolitan areas, has been and remains an (increasingly) serious, in fact threatening, problem that calls for drastic and ground-breaking solutions. Traffic congestion causes excessive delays, substantial environmental pollution and reduced traffic safety. The cost of road traffic congestion in Europe exceeds € 120 billion per year, without accounting for the excess environmental pollution and the cost of traffic accidents, the latter being some four times higher. Similar figures apply in the case of U.S.A. traffic congestion and accident costs. Conventional traffic management measures are valuable [1,2], but not sufficient to address the heavily congested traffic conditions, which must be addressed in a more comprehensive way that exploits gradually emerging and future ground-breaking new capabilities of vehicles and the infrastructure.

During the last decade, there has been an enormous effort by the automobile industry, as well as by numerous research institutions to develop and deploy a variety of Vehicle Automation and Communication Systems (VACS) that will revolutionize the capabilities of individual vehicles. VACS may be distinguished in: Vehicle Automation Systems ranging from relatively weak driver support to highly or fully automated driving; and **Vehicle Communication Systems** enabling V2V (vehicle-to-vehicle) and V2I (vehicle-to-infrastructure) communication. Some low-automation VACS are already available in the market, such as ACC (Adaptive Cruise Control), which automatically controls the vehicle speed according to the desired speed selected by the driver; or adjusts the distance in case of a slower front vehicle. Moreover, numerous companies and research institutions have been developing and testing in real traffic conditions high-automation or virtually driverless autonomous vehicles that monitor their environment and make sensible decisions not only about car-following, but also about lane changing [3]. There is a variety of concepts employed for their movement strategies, ranging from AI (Artificial Intelligence) to optimal control methods. It should be noted that the relatively high-risk task of lane changing is particularly challenging, both methodologically and practically [4-6].

This paper launches the TrafficFluid concept, which is a novel paradigm for vehicular traffic, applicable at high levels of vehicle automation and communication and high penetration rates, as expected to prevail in the not-too-far future. Although we may lower the requirements eventually, we assume for now that vehicles communicate with each other (V2V) and with the infrastructure (V2I) at sufficient frequency, distance and bandwidth; and drive automatically, based on own sensors, communications and appropriate movement control strategy. Other than that, vehicles may be of various types (e.g. electric or with internal combustion engine) and sizes and may have a variety of desired (or allowed) maximum speeds and accelerations. Given these, the TrafficFluid concept is based on the following two combined principles:

1. Lane-free traffic: Vehicles are not bound to fixed traffic lanes, as in conventional traffic, but may drive anywhere on the 2-D surface of the road, see Fig. 1.
2. Nudging: Vehicles communicate their presence to other vehicles in front of them (or are sensed by them), and this may exert a "nudging" effect on the vehicles in front (under circumstances and to an extent to be specified later), i.e. vehicles in front may experience (apply) a pushing force in the direction of the line connecting the centers of the nudging vehicle and the nudged vehicle in front. Figure 2 illustrates a possible instance of resulting behavior. Figures 1 and 2 are snapshots from a preliminary microscopic TrafficFluid simulator; the % marked on each vehicle reflects its current speed as a percentage of its desired speed. Yellow vehicles drive currently with lower speed than desired due to hindering slower vehicles in front of them, which are therefore nudged; while blue vehicles have a current speed equal to the desired speed or higher, the latter in case they are nudged by vehicles behind them. In Fig. 2, the yellow vehicle has a higher desired speed than the two trucks on its left and right, therefore it nudges them aside (on the lane-free road), so as to pass between them and accelerate to its desired speed (thus becoming blue).

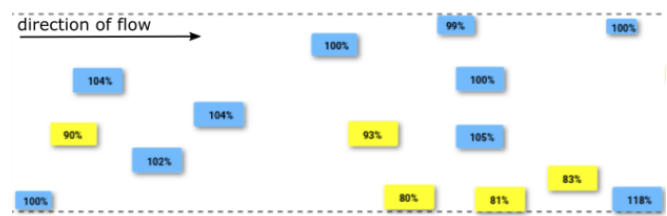

**Fig. 1.** Lane-free traffic.

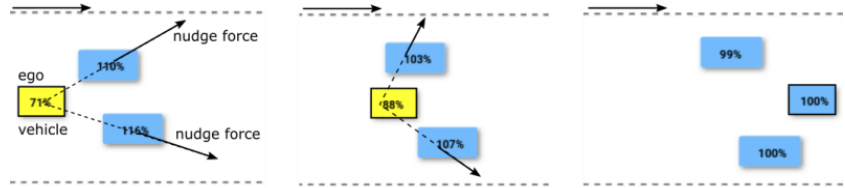

**Fig. 2.** An instance of the nudge-effect: The yellow vehicle nudges the slower trucks aside to pass.

## 2. Results

For a quick verification and demonstration of the outlined TrafficFluid concept, an ad-hoc model and microscopic simulator for the vehicle movement on a lane-free road was developed. The simulated vehicles are passenger cars (no trucks); they are randomly selected from 6 pre-specified vehicle-dimension classes; and have random desired speeds (within the pre-specified range [25, 35] m/s). All vehicles employ an identical movement strategy while driving on a circular road, whose 2-D surface has a length of 1 km and width of 10.2 m; this road width would barely suffice for 3 conventional motorway lanes.

The vehicle movement strategy is based on an "artificial forces" approach, whereby the longitudinal and lateral forces determine the corresponding vehicle acceleration in two dimensions. There are three 2-D forces acting in each direction (longitudinal and lateral). First, the target-speed force (positive or negative) depends on the deviation of the current vehicle speed from its desired speed; the latter being zero in the lateral direction. Second, each vehicle generates repulsive forces, fading with distance, that are applied to vehicles behind it and are introduced to avoid collisions. Third, each vehicle generates nudging forces, fading with distance, that are applied to vehicles ahead of it. Weighting parameters are used to adjust the impact of each force. After calculation of the forces, a bounding mechanism may clip them before they are used as vehicle accelerations; bounding aims at respecting various technical restrictions, in particular also the respect of the lateral road boundaries. The ad-hoc model is factually crash-free, i.e. no vehicle crashes were observed in all reported results, but we intend to derive more efficient and provably safe strategies in the future.

To assess and demonstrate some features of the TrafficFluid concept, the outlined simulation environment was used in a number of experiments. Specifically, four series of simulations were carried out, each series being summarized in a corresponding stationary flow (veh/h) versus density (veh/km) diagram, which is known as the Fundamental Diagram (FD), see Fig. 3. For each simulation run, the addressed number of vehicles (corresponding to a density value) are scattered roughly homogeneously on the road surface, whereby lateral vehicle positioning is closer to the left-hand boundary of the road for vehicles with higher desired speeds. All vehicles start with zero speed and accelerate eventually according to the vehicle movement strategy. After a transition period, the emerging traffic flow stabilizes around a stationary value, which is the value marked on the FD for the corresponding density. Four FDs are displayed in Fig, 3; in the first one, nudging forces are switched off, while two more FDs were produced with weak and stronger nudging, respectively. Finally, conditions on the fourth FD are identical as in the third, but the road width has been enlarged by 1.7 m, which corresponds to half-width of a conventional motorway lane, These summarized simulation results demonstrate that:

- In all cases we obtain the characteristic inverse-U shape of a conventional FD.
- The achieved flows and capacity without nudging are much higher than what is usually observed on a conventional three-lane motorway, something that is attributed mainly to the lane-free traffic character.
- Nudging increases the flows and the capacity, as well as the critical density (at which the highest flow occurs).

- The incremental road widening (by half "lane") leads to further increases of the flows, the capacity, the critical density and the maximum density (at which flow and speed return to zero).

The details of the movement control strategy, the simulations and the obtained results are provided in Section 4.

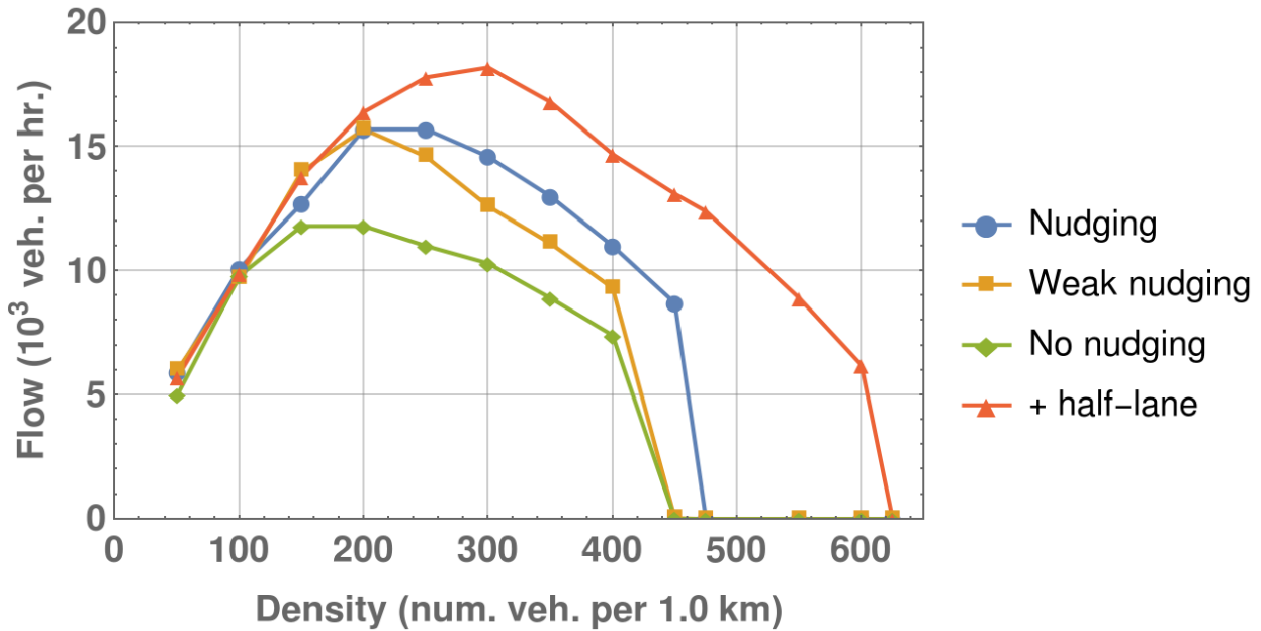

**Fig. 3.** Emerging flow-density curves (Fundamental Diagrams) for various simulated scenarios.

## 3. Discussion

### 3.A. General Features

For most of human history, roads did not need lanes because of low-speed movements. However, when automobiles came into widespread use during the beginning of the 20$^{th}$ century, there was a need to separate opposite traffic directions via lane markings on roads and highways to reduce the risk of frontal collisions; while dashed lines, separating parallel lanes on the same traffic direction, were only introduced in the 1950s, along with the rules governing lane-changing. Parallel lanes increase the traffic safety in manual driving, because they simplify the driving task for the human driver; when driving on a lane, the driver needs to monitor only the distance and speed of the front car, with virtually no need to also monitor the vehicle's left, right and rear sides. On the other hand, when a driver wishes to change her driving lane, things become more complex and risky, as the driver needs to look for an available gap on the target lane and predict its evolution based on the observed speeds of multiple vehicles (and of her own), while watching at the same time for the distance to the front vehicle. The lane-changing task becomes even more risky in cases of massive lane changes due to a lane-drop or merging on-ramps or roads. Lane changes are responsible for 10% of all accidents [7]. In summary, unidirectional lanes are indispensable in manual driving conditions due to increased safety; on the other hand, the existence of lanes entails the need for lane changing, which is recognized as an accident-prone manoeuvre.

The lane width on American interstate highways is 3.7 m, while German Autobahnen feature a lane width of 3.5 – 3.75 m. Since a medium-size car has a width in the order of 1.8 m, and a truck is some 2.5 m wide, we conclude that the lateral occupancy on motorways may be only slightly higher than 50%. Thus, the carriageway capacity could be strongly increased, even if only a part of the void lateral space is used, as in lane-free traffic. This indeed happens (semi-legally) to some extent in several developing countries, notably in India, where saturation (capacity) flow at traffic lights has been observed to increase strongly for inhomogeneous traffic with low lane discipline [8]. On top of the static capacity loss due to the need for wide lanes on high-speed highways with manual driving, additional capacity loss occurs due to dynamic phenomena attributed to lane-changing manoeuvres. Specifically, lane changing on highways is a notorious cause for reduced capacity [7] due to increased space occupancy of the lane-changing vehicle; and for triggering traffic breakdown at critical traffic conditions. Such phenomena are even more pronounced and detrimental to safety and capacity at locations of increased lateral movements, such as converging or diverging motorways, on- and off-ramps and weaving sections, because of the abrupt and space-consuming lateral displacements required in lane-based traffic.

In a nutshell, unidirectional traffic lanes have emerged in the mid-20$^{th}$ century as a necessary measure for improving traffic safety, even at the expense of reducing the highway capacity. According to the TrafficFluid concept, it may soon be time for the highways, motorways, arterials, and, perhaps, even urban roads to return to their lane-free structure, regaining the lost capacity and also improving on traffic safety. This can be achieved in the era of high-level vehicle automation and connectivity, as there is no need to mimic (in fact there are good reasons to avoid mimicking) the human lane-based driving task. Vehicle sensors and communications enable a CAV to monitor continuously its close and even distant surroundings on a 360º base and make fast moving decisions. These superbly increased capabilities, compared to human driving, would allow for a CAV to "float" safely and efficiently in a stream of other, potentially cooperating, CAVs, based on appropriate movement strategies.

Vehicle movement strategies for CAVs are easier to design, safer and more efficient in a lane-free environment due to smooth 2-D vehicle movement, where accident-prone, hence conservative, laterally "discontinuous" displacements to other lanes become obsolete. In addition, front-back vehicle collisions occurring in manual lane-based driving, sometimes involving dozens of vehicles in a pileup, may cause more serious damage than their counterpart of side-side collisions that are more likely to occur in lane-free traffic.

With regard to the second TrafficFluid principle, nudging, let us first note the (perhaps not merely) verbal similarity with the "nudge theory" by Richard Thaler that earned him the 2017 Nobel Prize in Economics. Thaler introduced the concept of "nudging" people through subtle changes in government policies, such that they do things that are beneficial for them in the long term (e.g. saving money). Back to traffic, a major and indeed "sacred" principle in traffic flow theory is the property of anisotropy in macroscopic traffic flow models [9-12]. Macroscopic traffic flow theory started with the pioneering work [13], which was based on an analogy of (single-lane, crowded) vehicular traffic flow with water flow in open channels. Vehicular traffic exhibits indeed many similar qualitative features as gas in a pipe or water flow in an open channel; similarly to water flow, traffic states propagate as waves with a speed different than the fluid particles speed; and shock waves form when fast vehicles catch up with slower vehicles in front. On the other hand, there is a major difference between water or gas flow versus vehicular traffic flow, which is due to the fact that vehicle movement (by the action of the human driver) is determined virtually exclusively by the happenings downstream (essentially by the distance and speed of the vehicle in front), while vehicles behind have normally no impact on a human's driving behaviour. In contrast, water or gas flow particles may influence the state of other downstream particles, e.g. fast particles may be "pushing" slower particles ahead making them accelerate. The fact that drivers react only to front vehicles is referred to as the anisotropic property of traffic flow and has specific mathematical consequences, e.g. that traffic waves cannot propagate faster than vehicles [9].

TrafficFluid's nudge effect enables vehicular traffic flow to be deliberately conceived in a variety of possible ways, without the anisotropy restriction imposed by human driving, so as to satisfy appropriate design criteria, e.g. maximize the road capacity. Note that nudging is much less interesting if applied to lane-based traffic, where some local inter-vehicle interaction might have a local stabilizing effect or slightly facilitate a lane change of the following vehicle, but this is not comparable to a generalized nudging policy that alters the characteristics of individual vehicle movement and, more importantly, of the emerging traffic fluid in a predictable engineered way. Naturally, nudging must be appropriately designed and limited; for example, nudging may be designed to have no effect if the nudged vehicle has already exceeded its desired speed by a certain percentage; and, certainly, nudging should not jeopardize traffic safety under any circumstances. Thus, vehicle nudging, in combination with lane-free flow, provide an unprecedented possibility to design (rather than describe or model) the traffic flow characteristics in an optimal way, subject to constraints, but without the need to satisfy anisotropy or other conditions stemming from the era of human driving. In short, we have the problem of designing, for the first time since the automobile invention, the properties of the traffic flow as an *artificial fluid*, and this is indeed the overarching feature of the TrafficFluid concept.

It is worth noting that the basic prerequisites for a real implementation of the TrafficFluid concept are moderate. On the vehicle level, the required movement strategy is likely to be easier to design than strategies currently deployed in autonomous vehicles for lane-based driving (including lane changes). With regard to on-board sensors and connectivity (V2V and V2I), there are no essential requirements that would exceed current plans for CAVs. Finally, TrafficFluid does not call for unconventional or costly new features for the road infrastructure. Note also that the TrafficFluid concept leads to incremental capacity increases, as a result of incremental road widening, in contrast to the need to widen conventional roads by lane "quanta". Thus, limited road widening around problematic bottleneck areas (e.g. on-ramps or strong upgrade or curvature) may be sufficient to dissolve local capacity problems that are those triggering congestion in conventional traffic.

### 3.B. Related issues

As the TrafficFluid concept is original, there is, as far as we are aware, no technical literature addressing issues related to it. Nevertheless, it is worth pointing out some works that may be considered to relate to some extent to the proposed TrafficFluid environment.

Microscopic simulation has been established in the last decades as a prominent tool for various traffic and transportation tasks. However, the low lane discipline and high number of small-size vehicles, including motorcycles, that are encountered in several developing countries, render lane-based simulation less accurate to reflect the actual traffic conditions in those countries. Therefore, in the last few years, there have been a few microscopic modelling works, which proposed, using various approaches, models for heterogeneous traffic (comprising different vehicle classes with different sizes and characteristics) and also for lane-less traffic. Overviews of such works may be found in [14,15]. Clearly, a major difference of the present work to these modelling works is that they attempt to *describe* the driving behaviour of real vehicles and drivers; while for TrafficFluid we need to *design* opportune movement strategies for safe and efficient traffic flow. The most recent of these works, which comes closest to TrafficFluid subjects, is [16], where a microscopic model for lane-less traffic is proposed, validated with real traffic data and analysed with respect to its stability properties by use of consensus-seeking agents methods [17]. Heterogeneous or multi-class traffic has also been considered in various ways in macroscopic modelling, see [14,18]. A recent macroscopic modelling work considers PTWs (Powered Two-Wheelers), which filter between cars, have particular dynamics, and do not respect lane discipline, similarly to a fluid in a porous medium, PTWs "percolate" between cars depending on the gap between them [19]. These works are rather remote from our endeavours, though some of them might provide useful hints regarding the development of macroscopic TrafficFluid models.

Regarding nudging or, more generally, the possibility for vehicles to influence the driving behaviour of other vehicles downstream, references are even sparser. While designing ACC regulators, [20] proposed the idea of using not only sensor measurements for the front distance, as usual in ACC, but also rear sensor measurements to the vehicle behind, so as to improve the stability properties of the ACC system. Clearly, using measurements referring to the vehicle behind is an instance of downstream influence of that vehicle. This idea was taken over in several other ACC-design works [21,22]. Despite reflecting influence from upstream, these works focus on lane-based longitudinal inter-vehicle stability issues and are therefore of marginal interest for our concept.

[23] proposes a novel approach to macroscopically model and simulate 2-D (longitudinal and lane-changing) lane-based traffic flow using the concept of Smoothed Particle Hydrodynamics (SPH). To this end, they propose a number of macroscopic "virtual forces" applying to elementary flow particles, which represent small volumes of the traffic flow, not necessarily vehicles. The forces are selected to reflect driver behavior for modelling conventional traffic; but for CAV traffic, the forces may differ and, in fact, they may also incorporate forward influence of vehicles, which comes close to our nudge effect, albeit at a macroscopic level and for lane-based traffic. Numerical simulations indicate that the application of forward forces may increase the road capacity, but, due to the macroscopic nature of the approach, it appears difficult to judge on the microscopic vehicle-level implications for safety and convenience. Nevertheless, the approach bears interest for our concept.

Another area that bears similarities with this work and has expanded enormously in the last decade is crowd modelling and simulation, see the book [24] and the overviews [25-27]; a similar area involving, beyond pedestrians, also cyclists and vehicles is traffic modelling in shared spaces [28]. A popular approach, while modelling moving persons, is to apply potential fields (a concept stemming from robotics path planning [29]), called "social forces" around each person in the 2-D space. Social forces reflect a variety of possible person intentions and knowledge, infrastructure types and constraints. One such social force is a repulsive force applied by a circular field around the centre of each (circular) moving person; this repulsive force fades out with distance from the person's center in the 2-D space and is included in the modelling to prevent collisions with other persons. In case two persons collide, i.e. they touch each other, as for example in emergency or high-density situations, then special "pushing" forces apply, which act similarly as our nudging, albeit only in case of adjacent colliding persons. In summary, crowd modelling is similar to TrafficFluid in that it may contain instances of lane-free moving of persons along a bounded path, but it has also significant differences: (a) It is a modelling approach aiming at mimicking real movements, not a design procedure for safe and efficient traffic flow; and (b) it addresses situations quite different from high-speed vehicles driving on roads. Nevertheless, the crowd modelling area includes some elements that may be of interest for our concept.

Finally, we note the existence of works referring to "Artificial Transportation Systems", see e.g. [30], where, however, the term "artificial" reflects essentially a "simulated" transportation system, which represents and replaces the real system; while the term "artificial fluid" in this paper's title is used literally and actively, reflecting the endeavour to design, deliberately and purposefully, an engineered traffic flow system.

**3.C. Developments required**

The TrafficFluid concept calls for substantial investigations to understand implications, exploit opportunities and conceive a safe and efficient artificial fluid of traffic. Such investigations must address, among others, the following challenging subjects: Vehicle movement strategy design for various scenarios of connectivity (V2V and V2I); consideration of different vehicle types, including trucks, emergency vehicles and manually driven vehicles; the possibilities and impact of forming vehicle platoons within the lane-free environment; the impact of incidents and congestion; consideration of various road infrastructures (motorways, arterials and even road junctions) with laterally entering and exiting traffic; development of realistic simulators; emerging macroscopic traffic flow model development; possible traffic-responsive (i.e. depending on the prevailing traffic conditions) actions, at the vehicle or traffic levels. In short, the proposed TrafficFluid concept and related investigations address a novel traffic environment that must be designed from scratch. We expect that TrafficFluid will trigger new research by many capable research groups to address some of the outlined, as well as additional issues, so as to explore the potential benefits of the new traffic paradigm while addressing a major problem for modern society.

## 4. Modeling and Simulation Details

### 4.A. Vehicle movement strategy

Longitudinal (x-direction) and lateral (y-direction) accelerations $f_x(t)$ and $f_y(t)$ for each vehicle at time $t$ are computed via the following respective equations:

$$f_x(t) = f_x^{ts}(t) + c_x^{ca}(f_x^{rp}(t) + \gamma_x f_x^{ng}(t)) \tag{1}$$

$$f_y(t) = f_y^{ts}(t) + c_y^{ca}(f_y^{rp}(t) + \gamma_y f_y^{ng}(t)) \tag{2}$$

where, $c_x^{ca}, c_y^{ca}$ are scalar coefficients for the longitudinal and lateral, respectively, "forces" $f^{ts}, f^{rp}, f^{ng}$. The purpose of each force is briefly described as follows:
- target-speed force $f^{ts}$ strives for the vehicle to attain its desired longitudinal (lateral) target speed;
- repulsive force $f^{rp}$, due to vehicles ahead, aims at preventing collision with such vehicles;
- nudging force $f^{ng}$, due to vehicles behind, aims at facilitating advancement of faster vehicles behind.

Finally, the scalar coefficients $\gamma_x, \gamma_y \in [0,1]$ adjust the effect of nudging forces in relation to repulsive forces; enabling attenuation of nudging forces as needed.

### 4.B. Target-speed force

Each vehicle is associated with a corresponding desired (non-zero) longitudinal speed $v_d$ and a zero lateral desired speed. When a vehicle is moving below, at, or above its longitudinal (lateral) desired speed, $f_x^{ts}$ ($f_y^{ts}$) becomes positive, zero, or negative, accordingly. More specifically, we have the following relations (illustrated in Fig. 4)

$$f_x^{ts}(t) = -\mathrm{erf}[v_x(t) - v_d] \tag{3}$$

$$f_y^{ts}(t) = -\mathrm{erf}[v_y(t)] \tag{4}$$

where erf is the well-known Gauss Error Function; used for its suitable sigmoid shape.

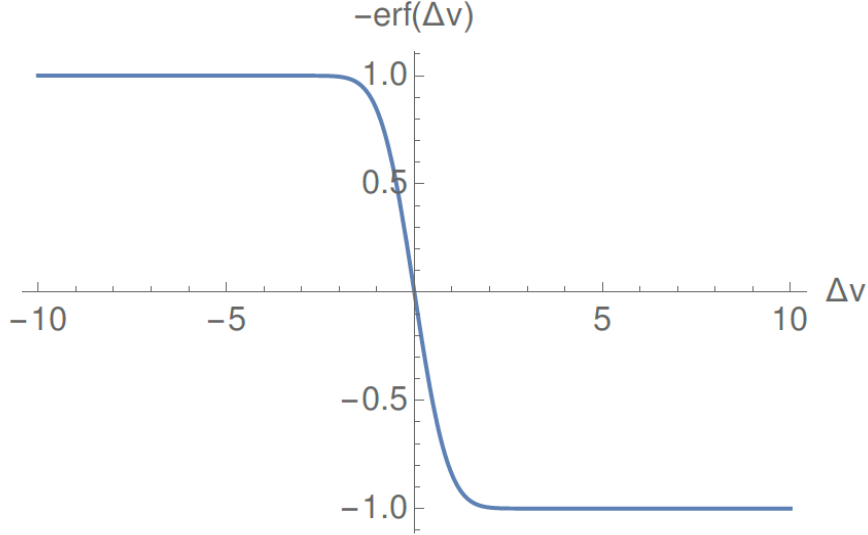

**Fig. 4:** The implementation of target-speed forces relies on the Error Function $\text{erf}$, providing deceleration or acceleration as needed to address any deviation from a desired speed.

### 4.C, Repulsive and nudging forces

Consider two vehicles $i$ and $j$, such that $i$ is upstream of $j$ and their longitudinal distance is less than $\Delta$. Then, vehicle $j$ exerts a repulsive force onto $i$; and vehicle $i$ exerts a nudging force onto $j$. The two forces have the same magnitude, but the nudging force may be eventually moderated by use of weights $\gamma_x, \gamma_y$ smaller than 1, as in (5) and (6). Both forces are applied along the line connecting both vehicle centers, but in opposite directions, as the repulsive force is applied to the upstream vehicle $i$ due to the presence of downstream vehicle $j$; while the nudging force is applied to the downstream vehicle $j$ due to the presence of upstream vehicle $i$.

To illustrate repulsion, from the perspective of upstream vehicle $i$, the downstream vehicle $j$ may be considered to be surrounded by a potential field or "aura", so that the positioning of $i$ within $j$'s aura determines the magnitude of the repulsive force. The extent and shape of the aura surrounding $j$ from the perspective of $i$ depends on: 1) the location of $j$; 2) the physical dimensions of vehicles $i$ and $j$; and 3) the lateral and longitudinal speeds of both $i$ and $j$. Hence, there is no single potential field surrounding a vehicle, as its shape depends on the perspective of an "observing" vehicle.

For illustrative purposes, let us consider the example contour plot presented in Fig. 5A. An upstream vehicle $i$ is moving with longitudinal speed $v_{i;x} = 25$ m/s, centered at position $(0,4)$ (in m). Downstream vehicle $j$ (blue; to its front-left) is centered at $(20, 5.7)$ and moving with longitudinal speed $v_{j;x} = 20$ m/s. Downstream vehicle $k$ (blue; to its front-right) is centered at $(25, 0.6)$ and moving with longitudinal speed $v_{k;x} = 20$ m/s. The lateral speed of all three vehicles is zero. Upstream vehicle $i$ is centered within the potential field created by its front-neighbors, which is the sum of the individual potential fields surrounding each neighbor. Clearly, only the potential field surrounding $j$ affects $i$, which receives a repulsion force. Let this force be denoted as $f_j^{rp} = (f_{j;x}^{rp}, f_{j;y}^{rp})$, consisting of a longitudinal component $f_{j;x}^{rp}$ and a lateral component $f_{j;y}^{rp}$. This way, upstream vehicle $i$ is "pressured", as far as the repulsive forces are concerned, to move away from $j$ in both the lateral and longitudinal direction.

A different situation is illustrated in Fig. 5B. Here, all vehicles are positioned as above, but now vehicle $j$ is moving as fast as vehicle $i$, i.e. at 25 $m/s$. Since $i$ is not approaching vehicle $j$, the aura of $j$ (from the perspective of $i$) is now inconsequential to $i$ (i.e. no repulsion is acting from $j$ to $i$), as $i$ is not centered within it. But, in contrast to the previous case, now vehicle $k$ is moving laterally towards $i$ at a lateral speed $v_{k;y} = 4\ m/s$. Due to its non-zero lateral speed, the aura enclosing $k$ extends laterally (towards the direction of lateral movement). Vehicle $i$ is now within $k$'s aura and receives a corresponding repulsive force. In effect, $i$ now perceives $k$ as a potential obstacle to be avoided, with the resulting repulsive force influencing $i$'s movement away from $k$, in both the lateral and longitudinal direction.

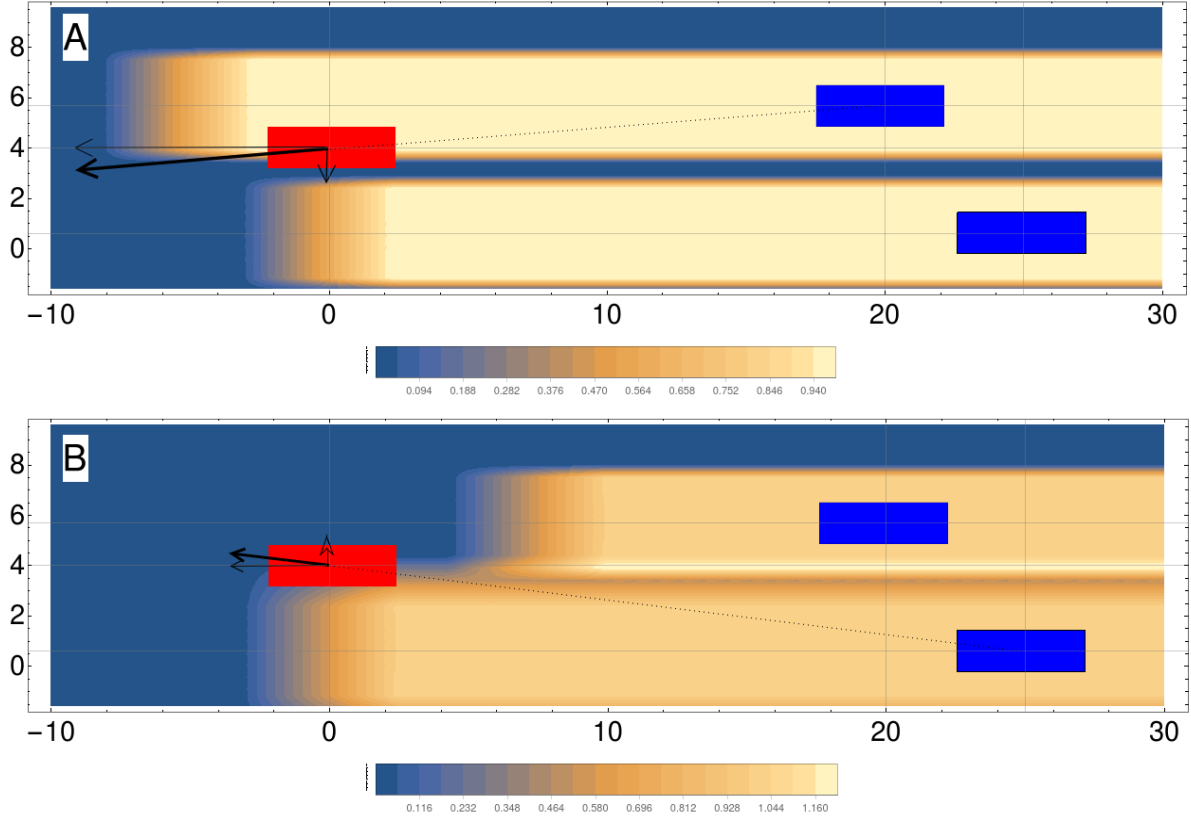

**Fig. 5:** A: Upstream vehicle $i$ (red; positioned at $(0,4)$) inside the potential field surrounding downstream vehicle $j$ (blue; positioned at $(20,5.7)$) receiving an appropriate backwards repulsive force $f_j^{rp} = (f_{j;x}^{rp}, f_{j;y}^{rp})$. B: Vehicle $i$ now positioned within the potential field surrounding downstream vehicle $k$ (blue; positioned at $(25, 0.6)$), now receiving an appropriate backwards repulsive force $f_k^{rp} = (f_{k;x}^{rp}, f_{k;y}^{rp})$.

Returning to the couple of upstream vehicle $i$ and downstream vehicle $j$, the downstream vehicle $j$ receives a nudging force from upstream vehicle $i$. The magnitude of this nudging force $f_i^{ng} = (f_{i;x}^{ng}, f_{i;y}^{ng})$ exerted onto $j$ is equal to the opposite-directional repulsive force and may also be illustrated by means of an aura surrounding $i$, wherein downstream vehicle $j$ is located.

The potential function describing the aura surrounding a vehicle from the perspective of another vehicle, and hence the magnitude (which ranges within $[0,1]$) of the repulsive and nudging forces for a pair of vehicles, is described next in more detail.

## 4.D. The potential function

Consider the following function of two-dimensional coordinates $(x, y)$, based on which we shall define the potential field of a downstream vehicle $j$, from the perspective of an upstream vehicle $i$.

$$\Pi(x, y, L_1, L, L_2, W_1, W, W_2) := H(x, L_1, L, L_2) \cdot H(y, W_1, W, W_2) \tag{7}$$

where

$$H(\delta, d_{rise}, d_{flat}, d_{fall}) := \max\left\{0, \min\left\{1, 1 - \frac{\delta - d_{flat}}{d_{fall}}, \frac{d_{flat} + \delta + d_{rise}}{d_{rise}}\right\}\right\} \tag{8}$$

The shape of this function in the $(x, y)$-space of the road surface depends on the non-negative arguments $L_1, L, L_2$ and $W_1, W, W_2$, which determine the potential field in the longitudinal and lateral direction, respectively. An impression of $\Pi$ is given in Fig. 6. Effectively, for fixed $y = 0$, $\Pi$ rises linearly from 0 to 1 within interval $x \in [-L_1 - L, -L]$; remains constant at value 1 within $x \in [-L, L]$; and drops linearly to 0 within $x \in [L, L + L_2]$. It behaves similarly in the lateral direction for $x = 0$ according to arguments $W_1, W, W_2$.

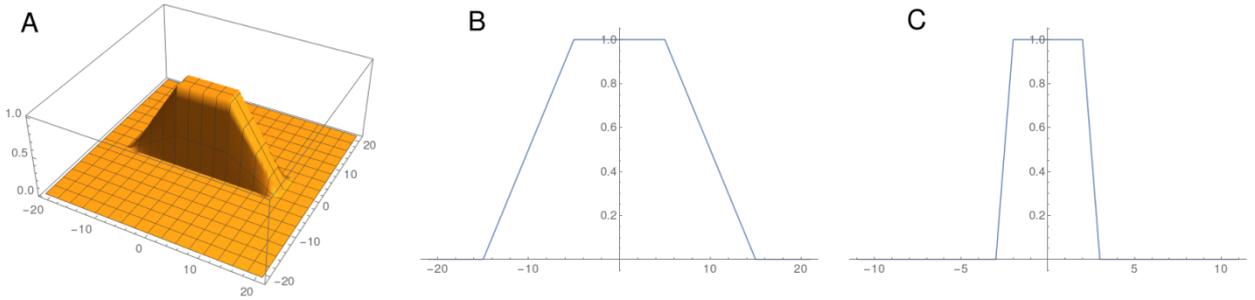

**Fig. 6:** A: 3D depiction of function $\Pi$ with $L_1 = L_2 = 10, L = 5$ and $W = 2, W_1 = W_2 = 1$. B: Cross-section for $y = 0, x \in [-25,25]$. C: Cross-section for $x = 0, y \in [-10,10]$.

From the perspective of $i$, vehicle $j$ is surrounded by a potential field defined as follows

$$\Pi\left(x - x_j, y - y_j, \frac{1}{2}L_x, \frac{1}{2}L_x + D_x, \frac{1}{2}L_x, L_y, D_y, L_y\right) \tag{9}$$

and the upstream vehicle $i$ receives, due to downstream vehicle $j$, a repulsion force $f_j^{rp}$, with magnitude $|f_j^{rp}| = \Pi \in [0,1]$, for $x = x_i, y = y_i$.

Term $L_x$ is chosen to be proportional to the longitudinal speed of vehicle $i$, so that the repulsive force is applied early enough to vehicles approaching fast from upstream; $D_x$ is equal to sum of half-lengths of the two vehicles, $0.5(l_i + l_j)$, plus a term accounting for the speed difference between the two vehicles, again for early repulsion of fast approaching upstream vehicles. In the lateral direction, since any two vehicles may be moving towards each other, $L_y$ is chosen to be proportional to the absolute value of the speed difference between the two vehicles $i$ and $j$; while term $D_y$ equals the sum of half-widths $0.5(w_i + w_j)$. This concept is illustrated in Fig. 7. More specifically, we have:

- $L_x = v_{i;x}T^x$ where $T^x$ is a parameter reminiscent of a longitudinal time-gap;
- $D_x = \frac{1}{2}\frac{\max\{0, v_{i;x} - \omega v_{j;x}\}^2}{c_x^{ca}} + \frac{1}{2}(l_i + l_j)$, where $c_x^{ca}$ is a parameter presented earlier;
- $L_y = |v_{i;y} - v_{j;y}|T^y$, where $T^y$ is reminiscent of a lateral time-gap, which, however, refers to the absolute value of the lateral speed differential; and
- $D_y = \frac{1}{2}(w_i + w_j)$.

The scalar parameter $\omega \in [0,1]$ is multiplied with $v_{j;x}$, so as to allow to adjust the extent to which $j$ is treated as a stationary obstacle; thus, for $\omega = 0$, downstream vehicle $j$ is seen as a stationary obstacle from the point-of-view of upstream vehicle $i$; while, for increasing values of $\omega$, vehicle $j$ is seen as moving accordingly slower. This parameter was introduced to impact the collision avoidance behavior of the strategy.

In the nudging case, the downstream vehicle $j$ receives, due to upstream vehicle $i$, a nudging force with an equal magnitude as above.

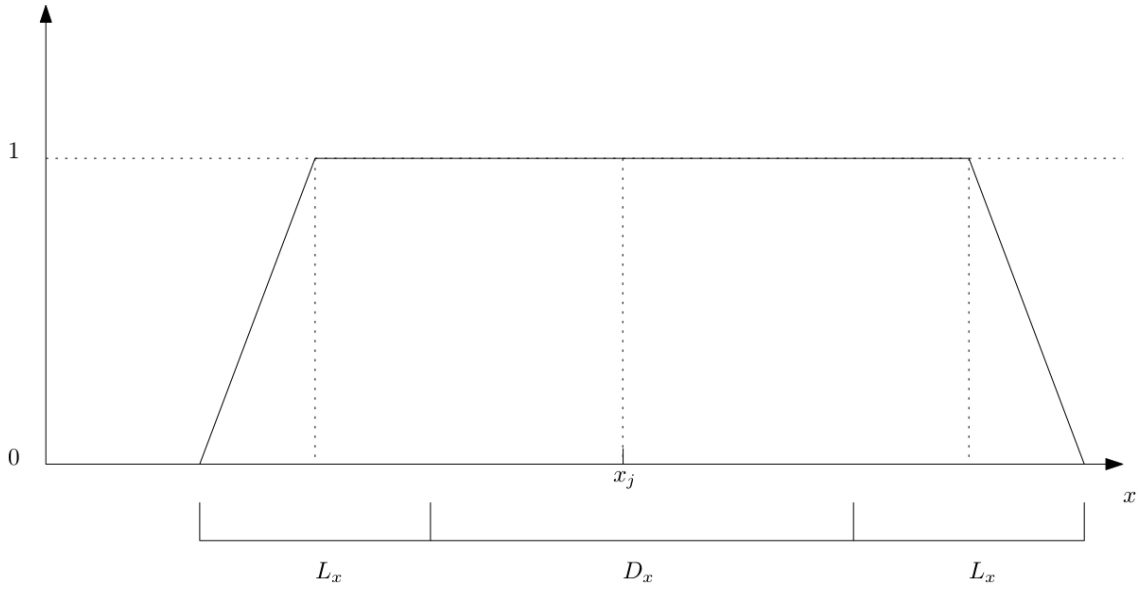

**Fig. 7:** Cross-section of potential function, illustrating the effect of parameters $L_x$ and $D_x$.

### 4.E. Accumulating repulsion and nudging forces

The repulsive force $f^{rp}$ acting on a reference vehicle $i$ equals the sum of individual repulsive forces $f_j^{rp}$ for every $j \in S^{rp}$, Where, for every reference vehicle $i$, the set $S^{rp}$ is defined as containing at most $n^{rp}$ downstream vehicles; namely those with the strongest repulsive influence within the forward distance $\Delta$. Similarly, set $S^{ng}$ is defined to contain at most $n^{ng}$ upstream neighbors; namely those with the strongest nudging influence over $i$ within the backward distance $\Delta$; and the nudging force $f^{ng}$ acting on a reference vehicle $i$ equals the sum of individual nudging forces $f_j^{ng}$ for every $j \in S^{ng}$.

### 4.F. Bounds

After $f_x$ and $f_y$ have been produced as a weighted sum of target-speed, repulsion and nudging forces, a constraining mechanism, described here, may clip them, so as to respect certain bounds, before they are used as accelerations.

Technical bounds: Longitudinal and lateral acceleration are bounded to lie within specified respective ranges $f_x \in [u_x^{b-}, u_x^{b+}]$ and $|f_y| \leq u_y^b$.

Road-boundaries: Consider a vehicle at time $t$, located at distance $d(t)$ from the road boundary and driving towards it with (lateral) speed $v(t) \geq 0$. It can be shown that collision with the boundary is avoidable if and only if $v(t) \leq \sqrt{-2(-u_y^b)d(t)}$, where $-u_y^b$ is the maximum applicable lateral deceleration. This speed bound, which we call "escape velocity bound", tends to zero, as $d(t)$ tends to zero (at a rate that depends on $u_y^b$). Thus, having the lateral velocity of a vehicle never exceeding the bound corresponding to its distance to the road boundary, is a necessary and sufficient condition for ensuring that no vehicle exits the road. In addition, it can be shown that having lateral acceleration at time-step $t$ not exceeding the following bound

$$\overline{U_t} = \frac{1}{2}\left[-u_y^b - 2Tv(t) + \sqrt{-u_y^b - \frac{4u_y^b v(t)}{T} + \frac{8(-u_y^b)[Tv(t) - d(t)]}{T^2}}\right] \quad (10)$$

ensures that lateral velocity $v(t+T)$ (at the beginning of the next time-step) will not violate the bound corresponding to its resulting distance to the road boundary.

Using eq. (8) with the current distance $d(t)$ and lateral speed $v(t)$ to each boundary of the road, a lower and an upper bound is derived and used to clip lateral acceleration $f_y$ at each time-step.

Longitudinal car-following bounds: These bounds were found to suppress some few vehicle collisions that were observed without its application. As explained earlier, every vehicle $i$ compiles a set $S^{rp}$ of downstream vehicles with the greatest repulsive impact. For each $j \in S^{rp}$, a corresponding upper bound $\zeta_j$ for the longitudinal acceleration is computed, as follows:

$$\zeta_j(t) = (1-\lambda)\frac{v_d - v_x(t)}{T} + \lambda\frac{\hat{v} - v_x(t)}{T} \quad (11)$$

Let $\lambda \in [0,1]$ indicate the extent to which $j$ is a potential downstream obstacle; i.e. $\lambda = 1$ when $j$ is dangerously near and laterally aligned; and $\lambda = 0$ when $j$ is far ahead or laterally strongly misaligned. Depending on $\lambda$, $\zeta_j(t)$ ranges from:
- the acceleration required to reach desired speed within one time-step (first term in (9));
- to the deceleration required to adjust the longitudinal speed to $\hat{v}$, which is defined as

$$\hat{v} = \min\left\{\frac{\Delta x}{T^x}, v_{j;x}(t)\right\} \quad (12)$$

Thus, with $\Delta x$ denoting the longitudinal distance to $j$, $\hat{v}$ is the minimum between:
- the speed corresponding to the time-gap-like parameter $T^x$;
- the longitudinal speed of the downstream vehicle $j$.

We use $\lambda = \left(f_{j;x}^{rp}\right)^\eta$ with $\eta = 0.25$, i.e. the extent to which a downstream vehicle $j$ is treated as an obstacle to be avoided is determined by where, within its repulsive aura, vehicle $i$ is located. As a result, $\lambda$ not only takes into account relative positioning, but also relative speeds, i.e. $\lambda$ may take high values even if $j$ is laterally misaligned but moving towards vehicle $i$, as in Fig. 5B.

Among all vehicles $j \in S^{rp}$, the vehicle $j^*$ with the tightest bound may be considered as a "leader", whose bound $\zeta_{j^*}$ is imposed as an upper bound for the longitudinal acceleration of vehicle $i$.

Other bounds: As a consequence of nudging forces, a vehicle may exceed its desired speed. Moreover, due to repulsive forces, a vehicle might move backwards. In order to prevent both of these cases, longitudinal acceleration $f_x$ is subjected to the following constraints

$$\frac{-v_x}{T} \leq f_x \leq \frac{(1+\alpha)v_d - v_x}{T} \tag{13}$$

effectively preventing a vehicle's longitudinal speed from exceeding $(1+\alpha)v_d$, with $\alpha > 0$, or from moving backwards.

Finally, lateral acceleration $f_y$ is subjected to the following constraints

$$\frac{-(\beta v_x - v_y)}{T} \leq f_y \leq \frac{\beta v_x - v_y}{T} \tag{14}$$

to ensure lateral speed will never exceed a fraction $\beta \in (0,1]$ of the current longitudinal speed, in either direction. In our simulations we are using $\alpha = 0.2$ and $\beta = 0.3$.

## 4.G. Experiment design

A simulation-based experiment was designed by use of the described ad-hoc vehicle movement strategy, in order to highlight and demonstrate some basic features of the TrafficFluid concept. To this end, we produced four fundamental diagrams (FDs), i.e. flow-versus-density stationary relations, for four corresponding scenarios of vehicle movement, as follows:
- No nudging forces applied ($\gamma_x = \gamma_y = 0$);
- Nominal (full) nudging forces applied ($\gamma_x = 0.9, \gamma_y = 1$);
- Moderate nudging forces applied ($\gamma_x = 0.45, \gamma_y = 0.5$); and
- Nominal nudging forces applied and incremental widening of the road.

Each FD was obtained by simulating the movement of $n$ vehicles (for multiple $n$), governed by the above movement strategy, on a (two-dimensional) ring-road, which is 1 km long and 10.2 m wide (corresponding to three narrow motorway lanes, each 3.4 m wide) for the first three scenarios; while the road was widened by 1.7 m (half conventional lane width) for the fourth scenario. The right-hand boundary of the road is defined by $y = 0$; and the left-hand boundary of the road is defined by $y = 10.2$; or 11.9 for the fourth scenario.

Let $(x(t), y(t))$ denote the position of (the center of) a vehicle at time $t$ during the simulation. Initially, at $t = 0$, each vehicle is placed at some location $(x(0), y(0))$, such that all $n$ vehicles are randomly and quasi-uniformly distributed on the 2-D road surface. To this end, we start with the lateral distribution and assign the vehicles to three zones (corresponding to "lanes"), with small lateral alignment imperfections. More specifically, the $n$ vehicles are equally divided into three (as many as lane-widths) parallel zones $l = 1, 2, 3$; and each zone is centered at $y_l = 1.7 + 1.7(l-1)$ m from the right-hand road boundary. Each vehicle allocated to zone $l$ is laterally centered at $y(0) = y_l + \varphi$, with $\varphi$ uniformly distributed within $[-0.5, +0.5]$. Eventually, the vehicles in each zone are longitudinally distributed at random (uniformly) along the 1 km of road length, and this assigns to them also a longitudinal initial coordinate $x(0)$.

The initial longitudinal (lateral) speed $v_x(0)$ ($v_y(0)$) of each vehicle is zero. The desired longitudinal speed $v_d$ assigned to a vehicle is a function of its initial lateral position $y(0)$, as follows

$$v_d = 25 + (35 - 25)\frac{y(0)}{10.2}$$

Thus, the closer a vehicle is positioned to the left-hand (right-hand) road boundary, the closer its desired speed is to $35\ m/s$ ($25\ m/s$). The dimensions of each of the $n$ vehicles placed on the ring-road are determined by choosing randomly (with the uniform distribution) one out of the six "dimension classes" reported in **Table 1**.

**Table 1:** The different dimension classes of vehicles used in the simulations.

|  | Class 1 | Class 2 | Class 3 | Class 4 | Class 5 | Class 6 |
|---|---|---|---|---|---|---|
| Length $l_i$ (m) | 3.20 | 3.90 | 4.25 | 4.55 | 4.60 | 5.15 |
| Width $w_i$ (m) | 1.60 | 1.70 | 1.80 | 1.82 | 1.77 | 1.84 |

To obtain each FD, we run a series of simulations employing the vehicle movement strategy settings corresponding to the specific FD scenario. Each simulation in such a series produces a point of the corresponding FD, i.e. it produces a density (veh/km) and a flow (veh/h) value. Specifically, each simulation series considers the presence of $n = 50, 100, 150, \ldots, 450$ vehicles on the 1 km long road, something that determines immediately the respective density (veh/km) values for each simulation of the series (note that also higher $n$ values are considered in the fourth scenario due to a wider road). In order to also obtain a flow value for each $n$, a simulation is run for $K = 6000$ time-steps of length $T = 0.2\ s$, corresponding to some 20 min of simulated time. At each time $t$, each vehicle is moved according to the longitudinal and lateral accelerations $f_x(t)$ and $f_x(t)$, respectively, deriving from the described vehicle movement strategy, based on the following kinematic equations for 2-D vehicle position and speed

$$x(t+T) = x(t) + v_x(t)T + 0.5\ f_x(t)T^2$$

$$y(t+T) = y(t) + v_y(t)T + 0.5\ f_y(t)T^2$$

$$v_x(t+T) = v_x(t) + T$$

$$v_y(t+T) = v_y(t) + f_y(t)T$$

for $t = 0, T, 2T, 3T, \ldots, (K-1)T$. Thus, at the beginning of each time-step $t$ of length $T$, each vehicle departs from position $(x(t), y(t))$ with longitudinal (lateral) speed $v_x(t)$ ($v_y(t)$); and, moving with constant longitudinal (lateral) acceleration $f_x(t)$ ($f_y(t)$), it reaches its updated state at time $t + T$. Starting with the described initial state for all vehicles at time 0, the behavior of the overall system reaches, after a transition period, a quasi-steady state in terms of the emerging macroscopic traffic states. The traffic flow (in veh/h) is measured at a specific road location and is averaged for the last 1500 time-steps of the simulation to produce the stationary traffic flow value corresponding to density $n$ in the specific FD.

For all produced FDs, the longitudinal time-gap-like parameter is set to $T^x = 0.65$ s, and the lateral one is set to $T^y = 0.25$ s. As constant bounds for longitudinal and lateral acceleration, we use $u_x^{b-} = -4.5\ m/s^2$, $u_x^{b+} = 2.0\ m/s^2$, and $u_y^b = 1.5 m/s^2$, respectively. Parameter $\omega$, which determines the extent to which any downstream vehicle is treated as a stationary obstacle, is set to $\omega = 0.3$. The maximum number of upstream (downstream) vehicles $n^{ng}$ ($n^{rp}$) taken into account in determining a nudging (repulsive) force exerted on a vehicle are set to $n^{ng} = 0, n^{rp} = 6$ for the "no nudging" case and $n^{ng} = 3, n^{rp} = 6$ for the rest of scenarios, which consider nudging.

**Acknowledgments:** We thank A.I. Delis and I.K. Nikolos for inspiring discussions on the TrafficFluid concept.

**Funding:** This research started in the final phase of, and received funding from the European Research Council [ERC grant number 321132] project TRAMAN21 (2013-2018); as well as from related matching funds by the Greek Ministry of Education, Research and Religious Affairs (General Secretariat for Research and Technology); the TrafficFluid concept will be developed in technical detail within the European Research Council [ERC grant number 833915] project TrafficFluid (2019-2024).

**References**

[1] Papageorgiou, M., Diakaki, C., Dinopoulou, V., Kotsialos, A., Wang, Y., 2003. Review of road traffic control strategies. *Proceedings of the IEEE* 91, 2043-2067.
[2] Papageorgiou, M., Ben-Akiva, M., Bottom, J., Bovy, P.H.L., Hoogendoorn, S.P., Hounsell, N.B., Kotsialos, A., McDonald, M., 2007. ITS and Traffic Management. In *Transportation (Handbooks in Operations Research and Management Science Vol. 14)*, C. Barnhart and G. Laporte, Editors, North-Holland (Elsevier), 715-774.
[3] Aeberhard, M., Rauch, S., Bahram, M., Tanzmeister, G., Thomas, J., Pilat, Y., Homm, F., Huber, W., Kaempchen, N., 2015. Experience, results and lessons learned from automated driving on Germany's highways, *IEEE Intelligent Transportation Systems Magazine* 7, 42-57.
[4] Ardelt, M., Coester, C., Kaempchen, N., 2012. Highly automated driving on freeways in real traffic using a probabilistic framework, *IEEE Transactions on Intelligent Transportation Systems* 13, 1576-1585.
[5] Kamal, M.A.S., Taguchi, S., Yoshimura, T., 2016. Efficient driving on multilane roads under a connected vehicle environment, *IEEE Transactions on Intelligent Transportation Systems* 17, 2541-2551.
[6] Makantasis, K., Papageorgiou, M. 2018. Motorway path planning for automated road vehicles based on optimal control methods. *Transportation Research Record*, vol. 2672 (19), pp. 112–123.
[7] Rahman, M., Chowdhury, M., Xie, Y., He, Y., 2013. Review of microscopic lane-changing models and future research opportunities. *IEEE Transactions Intelligent Transportation Systems* 14, 1942-1956.
[8] Radhakrishnan P., Mathew, T.V., 2011. Passenger car units and saturation flow models for highly heterogeneous traffic at urban signalised intersections. *Transportmetrica* 7, 141-162.
[9] Daganzo, C.F., 1994. The cell transmission model: a dynamic representation of highway traffic consistent with the hydrodynamic theory. *Transportation Research Part* B 28, 269-287.
[10] Papageorgiou, M., 1998. Some remarks on macroscopic traffic flow modelling. *Transportation Research Part A* 32, 323-329.
[11] Zhang, H.M., 2003. Anisotropic property revisited – does it hold in multi-lane traffic? *Transportation Research Part B* 37, 561-577.
[12] Helbing, D., Johansson, A.F., 2009. On the controversy around Daganzo's requiem for and Aw-Rascle's resurrection of second-order traffic flow models. *The European Physical Journal B* 69, 549-562.
[13] Lighthill, M.J., Whitham, G.B., 1955. On kinematic waves: II. A theory of traffic flow on long crowded roads. *Proceedings of Royal Society* 229, 317-345.
[14] Munigety, C.R., Mathew, T.V., 2016. Towards behavioral modeling of drivers in mixed traffic

conditions. *Transportation in Developing Economies* 2:6, 20 pp.

[15] Asaithambi, G. Kanagaraj, V., Toledo, T., 2016. Driving behaviors: Models and challenges for non-lane based mixed traffic. *Transportation in Developing Economies* 2:19, 16 pp.

[16] Mulla, A.K., Joshi, A., Chavan, R., Chakraborty, D., Manjunath, D., 2019. A microscopic model for lane-less traffic. *IEEE Transactions on Control of Network Systems* 6, 415-428.

[17] Ren, W., Atkins, E., 2007. Distributed multi-vehicle coordinated control via local information exchange, *International Journal of Robust and Nonlinear Control* 17, 1002–1033.

[18] Bhavathrathan, B., Mallikarjuna, C., 2012. Evolution of macroscopic models for modeling the heterogeneous traffic: an Indian perspective. *Transportation Letters* 4, 29-39.

[19] Gashaw, S.M., Goatin, P. and Harri, J., 2018. Modeling and analysis of mixed flow of cars and powered two wheelers. *Transportation Research Part C* 89, 148-167.

[20] Zhang, Y, Kosmatopoulos, E.B., Ioannou, P.A., Chien, C.C., 1999. Autonomous intelligent cruise control using front and back information for tight vehicle following maneuvers, *IEEE Transactions on Vehicular Technology* 48, 319-328.

[21] Hao, H., Barooah, P., 2013. Stability and robustness of large platoons of vehicles with double-integrator models and nearest neighbor interaction. *International Journal of Robust and Nonlinear Control* 23, 2097-2122.

[22] Liu, Y., Pan, C., Gao, H., Guo, G., 2017. Cooperative spacing control for interconnected vehicle systems with input delays. *IEEE Transactions on Vehicular Technology* 66, 10692-10704.

[23] Zhang, Y., Zhang, G., Fierro, R., Yang, Y., 2018. Force-driven traffic simulation for a future connected autonomous vehicle-enabled smart transportation system, *IEEE Transactions Intelligent Transportation Systems*, online, doi: 10.1109/TITS.2017.2787141.

[24] Pelechano, N., Allbeck, J.M., Badler, N.I., 2008. *Virtual Crowds: Methods, Simulation and Control*. Morgan & Claypool Publishers.

[25] Haghani, M., Sarvi, M., 2018. Crowd behaviour and motion: Empirical methods. *Transportation Research Part B* 107, 253-294.

[26] Chen, X., Treiber, M., Kanagaraj V., Li, H., 2018. Social force models for pedestrian traffic – state of the art. *Transport Reviews* 38, 625-653.

[27] Duives, D.C., Daamen, W., Hoogendoorn, S.P., 2013. State-of-the-art crowd motion simulation models. *Transportation Research Part C* 37, 193-209.

[28] Anvari, B., Bell, M.G.H., Sivakumar, A., Ochieng, W.Y., 2015. Modelling shared space users via rule-based social force model. *Transportation Research Part C* 51, 83-103.

[29] Hwang, Y.K., Ahuja, N., 1992. A potential field approach to path planning, *IEEE Transactions on Robotics and Automation* 8, 23-32.

[30] Wang, F.-Y., Tang, S., 2005. A framework for artificial transportation systems: From computer simulations to computational experiments, *Proceedings of the 8$^{th}$ International IEEE Conference on Intelligent Transportation Systems*, Vienna, Austria, September 13-16, 2005, 1130-1134.